\documentclass[aps,pre,twocolumn,groupedaddress,showpacs,floatfix]{revtex4-1}

\usepackage{graphicx}
\usepackage{amssymb}
\usepackage{nicefrac}
\usepackage{xcolor}
\usepackage{here}

\usepackage{times}

\newcommand{\vm}{v_{\mathrm{m}}}
\newcommand{\xm}{x_{\mathrm{m}}}

\newcommand{\ek}{\mathcal{E}_k}
\newcommand{\ep}{\mathcal{E}_p}
\newcommand{\ee}{\mathcal{E}}

\usepackage{hyperref}

\hypersetup{
  colorlinks=true,
citecolor=black,
linkcolor=black,
urlcolor=black
  }

\begin{document}

\title{Energy partition for anharmonic, undamped, classical oscillators}

\author{Micha{\l} Mandrysz}
\email{{michal.mandrysz@student.uj.edu.pl}} 
\affiliation{
Institute of Theorethical Physics, and Mark Kac Center for Complex Systems
Research, Jagiellonian University, ul. St. {\L}ojasiewicza 11,
30--348 Krak\'ow, Poland}
\author{Bart{\l}omiej Dybiec}
\email{{bartek@th.if.uj.edu.pl}} 
\affiliation{
Institute of Theorethical Physics, and Mark Kac Center for Complex Systems
Research, Jagiellonian University, ul. St. {\L}ojasiewicza 11,
30--348 Krak\'ow, Poland}

\date{\today}

\begin{abstract}
Using stochastic methods, general formulas for average kinetic and potential energies for anharmonic, undamped (frictionless), classical oscillators are derived.
It is demonstrated that for potentials of $|x|^\nu$ ($\nu>0$) type energies are equipartitioned for the harmonic potential only.
For potential wells weaker than parabolic potential energy dominates, while for potentials stronger than parabolic kinetic energy prevails.
Due to energy conservation, the variances of kinetic and potential energies are equal.
In the limiting case of the infinite rectangular potential well ($\nu\to\infty$) the whole energy is stored in the form of the kinetic energy and variances of energy distributions vanish.
\end{abstract}


\maketitle



\section{Introduction and motivation\label{sec:introduction}}

Classical mechanics \cite{landau1988theoretical,goldstein2002classical} is a well developed and established theory. 
Newtonian, Lagrangian and Hamiltonian methods can be effectively used to describe and solve various problems \cite{kotkin2013collection,taylor2005classical}.
Already examination of toy models can be very insightfull and beneficial.
These models provide us with valuable intuition which can be transferred to more complex setups or into very different fields.
Unfortunately, not all models can be solved exactly.
Even for conceptually very simple setups, e.g., classical (purely) anharmonic oscillator, one has to rely on numerical \cite{press1992} or approximate methods \cite{viswanathan1957theory} like perturbative methods \cite{saletan1998classical,mandal2005approximate}.
Various numerical methods have been developed in quantum \cite{hioe1975quantum,biswas1973eigenvalues,voros1983quartic,brizuela2014classical} and classical setups \cite{apostol2005anharmonic} both for purely anharmonic potentials \cite{banerjee1978anharmonic} or for parabolic potential with anharmonic addition \cite{kesarwani1981eigenvalues,pathak2002classical,dong2019exact,levai2019exact}.
These methods focus on construction of exact or approximate solutions, eigenvalues, energies of ground states but they generally do not consider the problem of energy partition.

Here, we study the problem of energy partitionning in undamped (frictionless), (purely) anharmonic, classical oscillators.
We solve the problem solely on the basis of the classical mechanics without relaying on tools known from quantum mechanics or statistical physics \cite{robinett1997average,liboff2003introductory,huang1963,semay2016quantum}.
First, we recall the classical harmonic oscillator as the fully solvable case. 
Next, we generalize the potential to the purely anharmonic oscillators.
Despite the fact that studied models typically are not traceable analytically, we provide exact formulas for the time averaged energy partitions. 
We demonstrate a general framework which allows for calculation of arbitrary moments of energy distributions.
Therefore, our approach is more explicit than the Virial theorem \cite{goldstein2002classical}, which provides the formula for the ratio of average energies.

The applied methodology is conceptually similar to \cite{robnik2006exact,andresas2014statistical} where energy distributions in 1D time dependent oscillators are calculated.
The system studied here is time independent and undamped (frictionless), consequently its energy is conserved, i.e., it is constant.
Similarly, like in \cite{robnik2006exact,andresas2014statistical} we use uniform distribution of initial conditions on the constant energy curve to calculate time averaged energies (instead of energy distributions).
Uniform distribution of initial conditions accompanied with the ergodic theory can be used to calculate energy partition for any single-well potential of $|x|^\nu$ ($\nu>0$) type.

\section{Model and results}

The harmonic oscillator \cite{goldstein2002classical}
\begin{equation}
 m \frac{d^2 x(t)}{d t^2}=-\kappa x (t)\;\;\;\;\;\; (\kappa>0)
 \label{eq:harmonic-oscillator}
\end{equation}
is a fundamental fully solvable model of classical mechanics, which is used as the approximation of many complex systems.
Formula~(\ref{eq:harmonic-oscillator}) describes the fully deterministic, conservative system which undergoes periodic motion.
The general solution of Eq.~(\ref{eq:harmonic-oscillator}) is given by
$
 x(t)=A \sin(\omega t + \delta),
 \label{eq:harmonic-oscillator-solution}
$
where $\omega=\sqrt{\kappa/m}$ is the frequency of the harmonic motion.
Two constants $A$ (amplitude) and $\delta$ (phase shift) are determined by the initial conditions.
The total instantaneous energy $\ee$ of the harmonic oscillator is partitioned between kinetic  $\ek=mv^2/2$ and potential energy $\ep=\kappa x^2/2$. Average energies satisfy
$
 \langle \ee \rangle = \ee = \langle \ek \rangle + \langle \ep \rangle.
$
Using the general solution of Eq.~(\ref{eq:harmonic-oscillator}) and the fact that the period of the motion is $T=2\pi/\omega$ one can calculate time averaged energies
\begin{equation}
 \langle \ek \rangle = \frac{1}{T}\int_0^T \frac{m}{2}\dot{x}^2(t) dt=\frac{m}{4}\omega^2 A^2=\frac{\kappa}{4}A^2=\frac{\ee}{2}
 \label{eq:ek-partition-ho}
\end{equation}
and
\begin{equation}
 \langle \ep \rangle = \frac{1}{T}\int_0^T \frac{\kappa}{2} x^2(t) dt=\frac{\kappa}{4}A^2=\frac{\ee}{2},
 \label{eq:ep-partition-ho}
\end{equation}
because $\ee=\kappa A^2/2$.
Consequently, for the harmonic oscillator, average kinetic and potential energies are equipartitioned, i.e., they are equal to each other and to half of the total energy.
The equipartition of average kinetic and potential energies takes place for the harmonic oscillator only.

For the general, anharmonic single-well potentials
\begin{equation}
    V(x)=\frac{\kappa}{\nu} |x|^\nu\;\;\;\;\;\;\;\;\;\;\;\;(\kappa>0,\nu>0).
    \label{eq:potential}
\end{equation}
the Newton equation~(\ref{eq:harmonic-oscillator}) reads
\begin{equation}
    m \frac{d^2 x(t)}{dt^2}=-V'(x)
    \label{eq:newton1d}
\end{equation}
Eq.~(\ref{eq:newton1d}) describes a periodic motion with the period $T$ given by
\begin{equation}
    T=\frac{2}{\nu}\sqrt{\frac{2\pi m}{\ee}}\left[ \frac{\nu\ee}{\kappa} \right]^{\frac{1}{\nu}} \frac{\Gamma\left( \frac{1}{\nu} \right)}{\Gamma\left( \frac{1}{2} + \frac{1}{\nu} \right)},
    \label{eq:period}
\end{equation}
where $\Gamma(\dots)$ is the Euler Gamma function, see Refs.~\cite{landau1988theoretical,amore2005exact}.
Due to the absence of damping the system described by Eq.~(\ref{eq:newton1d}) is conservative. 
The total instantaneous energy $\ee$ is constant and equal to
$
    \ee = \frac{1}{2}mv^2+V(x). 
$
It's exact value is determined by initial conditions, i.e., $v(0)$ and $x(0)$.
The motion in phase space is restricted to the closed constant energy curve.
The energy conservation can be used not only to calculate $T$, see Eq.~(\ref{eq:period}), but also to calculate the implicit dependence $t(x)$, which typically cannot be inverted \cite{viswanathan1957theory}.
Consequently, in order to calculate the energy partition, it is not possible to use the methodology applied in Eqs.~(\ref{eq:ek-partition-ho}) and (\ref{eq:ep-partition-ho}).
Here, using the probabilistic approach \cite{mandrysz2019deterministic}, we calculate the energy partition for any $V(x) \propto |x|^\nu$ ($\nu>0$) potential.

From a single trajectory, after long measurement time, it is possible to estimate the probability $p(x)$ of observing the particle in the neighborhood of $x$. 
Applying the correspondence principle \cite{liboff2003introductory,robinett1995quantum,robinett2002visualizing}, this probability is proportional to the time spent in the vicinity of $x$
\begin{eqnarray}
    p(x)dx  & = &  \frac{dt}{\frac{T}{2}} = \frac{2}{T}  \frac{dt}{dx}dx = \frac{2}{T}\frac{dx}{\frac{dx}{dt}} = \frac{2}{T}\frac{dx}{v},
    \label{eq:dtratio}
\end{eqnarray}
where $T$ is the period of the motion given by Eq.~(\ref{eq:period}).
In the denominator of Eq.~(\ref{eq:dtratio}) there is $\frac{T}{2}$ because the motion between two most distanced (turning) points ($\pm \xm$) lasts half of the period, i.e., $\frac{T}{2}$.
During the full period $T$ every point $x$ is visited twice with corresponding velocities $v(x)$ and $-v(x)$.
Therefore, in order to calculate $p(x)$ one needs to study the appropriate part of the full motion only.
From the energy conservation principle one can calculate the velocity
$
    v=\sqrt{\frac{2}{m}\left[ \ee- \frac{\kappa}{\nu}  |x|^\nu  \right]},
$
where $\ee=\frac{m}{2}v^2+\frac{\kappa}{\nu}|x|^\nu$ is the total energy.
The probability density $p(x)$ then reads:
\begin{equation}
    p(x)= \frac{2}{T}\frac{1}{\sqrt{\frac{2}{m}\left[ \ee-\frac{\kappa}{\nu}  |x|^\nu  \right]}}.
    \label{eq:px}
\end{equation}
The probability density $p(x)$ has maxima at points where the particle spends most time, see Eq.~(\ref{eq:dtratio}), i.e., at points where the velocity is minimal \cite{mandrysz2019deterministic}.
These points correspond to reversal points, i.e., to the points where the velocity smoothly changes its sign. These points are placed in the maximal distance $\pm \xm$ from the origin, see Eq.~(\ref{eq:px}).

The complementary $p(v)$ density  can be obtained from $p(x)$ by the transformation of variables
\begin{equation}
    p(v)=p(x(v)) \left| \frac{dx}{dv} \right|.
    \label{eq:trans}
\end{equation}
Using again the energy conservation one gets
\begin{equation}
    x= \left[ \frac{\nu}{\kappa } \left(  \ee -\frac{1}{2}mv^2  \right)  \right]^{\frac{1}{\nu}}
\end{equation}
and
\begin{equation}
    \frac{dx}{dv}=- \frac{m}{\kappa}\left[ \frac{\nu}{\kappa } \left(  \ee -\frac{1}{2}mv^2  \right)  \right]^{\frac{1}{\nu}-1} v.
    \label{eq:derivative}
\end{equation}
Finally, combining Eqs.~(\ref{eq:trans}) and (\ref{eq:derivative}) with the  energy conservation principle, one obtains
\begin{equation}
    p(v)=\frac{2m }{T\kappa } \left[ \frac{\nu}{\kappa } \left( \ee - \frac{1}{2}m v^2 \right)  \right]^{\frac{1}{\nu}-1}.
    \label{eq:pv}
\end{equation}
Maxima of the $p(v)$ density are located at points where the acceleration is minimal, i.e., the velocity is maximal \cite{mandrysz2019deterministic}.

Formulas for $p(x)$ and $p(v)$ densities, see Eqs.~(\ref{eq:px}) and (\ref{eq:pv}), can be used to derive main results of the current research, i.e., time averaged energies given by Eqs.~(\ref{eq:ek-partition}) -- (\ref{eq:ep-partition})
along with appropriate variances, see Eqs.~(\ref{eq:ek-var}) -- (\ref{eq:ep-var}), and formulas for moments of the arbitrary order which are given by formulas~(\ref{eq:cumv}) and~(\ref{eq:cumx}).
\begin{figure}[!ht]
    \centering    
    \includegraphics[width=0.97\columnwidth]{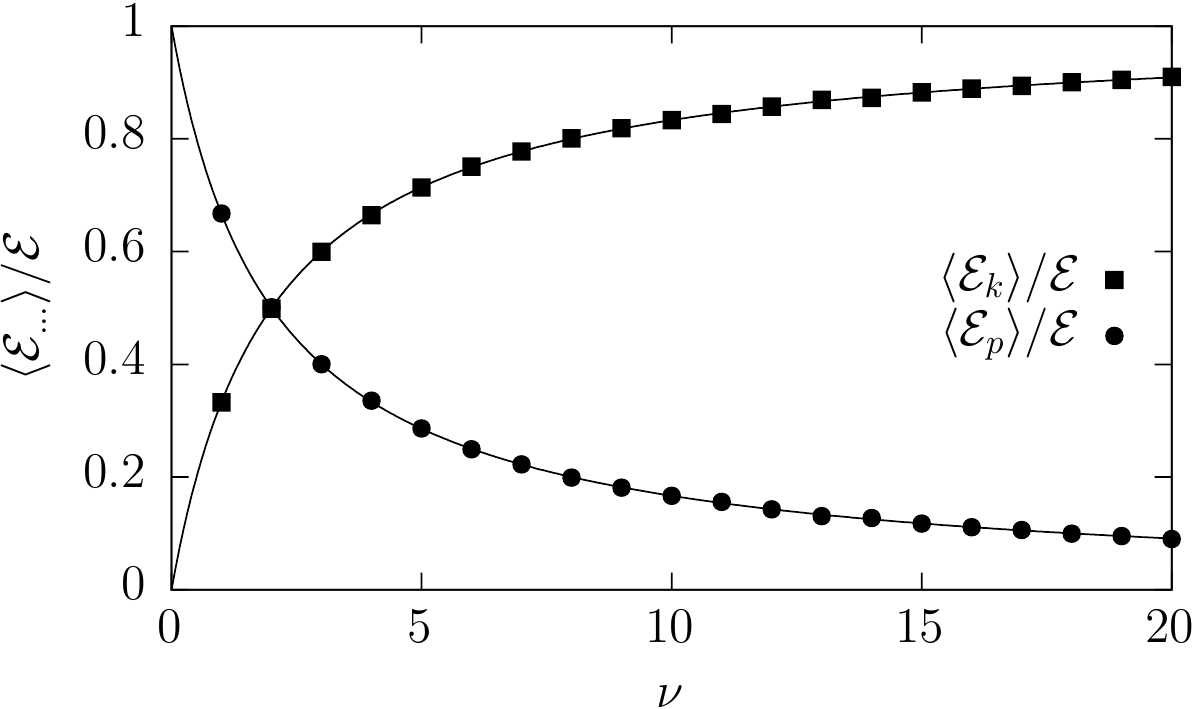} \\
        \caption{Rescaled average kinetic $\langle \ek \rangle/\ee$ (squares) and potential  $\langle \ep \rangle/\ee$ (circles) energies as a function of the exponent $\nu$ characterizing the steepness of the potential $V(x)$, see Eq.~(\ref{eq:potential}).
            Solid lines depicts the theoretical curve given by Eqs.~(\ref{eq:ek-partition}) and~(\ref{eq:ep-partition}) while points correspond to the results of computer simulations.
        }
    \label{fig:energy-partition}
\end{figure}

It can be verified that densities~(\ref{eq:px}) and (\ref{eq:pv}) are normalized on $[-\xm,\xm]$ and $[-\vm,\vm]$ intervals respectively, where the maximal allowed velocity $\vm$ and the maximal possible displacement $\xm$ satisfy the conditions
$\ee=\frac{m}{2}\vm^2$ and $\ee=\frac{\kappa}{\nu}|\xm|^\nu$.
Importantly, Eqs.~(\ref{eq:px}) and (\ref{eq:pv}) can be used to calculated the average kinetic and potential energies for any single-well potential $V(x) \propto |x|^\nu$
\begin{equation}
 \langle \ek \rangle = \int_{-\vm}^{\vm} p(v) \frac{m}{2} v^2 dv = \frac{\nu}{2+\nu} \ee
 \label{eq:ek-partition}
\end{equation}
and
\begin{equation}
 \langle \ep \rangle = \int_{-\xm}^{\xm} p(x) \frac{\kappa}{\nu} |x|^\nu dx = \frac{2}{2+\nu} \ee.
 \label{eq:ep-partition}
\end{equation}
Due to energy conservation, instantaneous energy $\ee$ is equal to the average energy $\langle \ee \rangle=\langle \ek \rangle + \langle \ep \rangle$.
The ratio of average kinetic and potential energies is equal to
\begin{equation}
 \frac{\langle \ek \rangle} { \langle \ep \rangle} = \frac{\nu}{2},
 \label{eq:ratio}
\end{equation}
which agrees with result of the Virial theorem \cite{goldstein2002classical}.
Formula~(\ref{eq:ratio}) demonstrates that energy equipartition is recorded for  $\nu=2$ only.
For $0<\nu<2$ the larger fraction of average energy is stored as the potential energy while for $\nu>2$ kinetic energy dominates, see Figure~\ref{fig:energy-partition}.
Furthermore, in the limit of $\nu\to\infty$, the potential $V(x)$, see Eq.~(\ref{eq:potential}), reduces to the infinite rectangular potential well.
In such a limit the whole energy is accumulated as the kinetic energy only.
Figure~\ref{fig:energy-partition} shows energy partitions as a function of the exponent $\nu$ characterizing the steepness of the potential $V(x)$, see Eq.~(\ref{eq:potential}).
Solid lines given by Eqs.~(\ref{eq:ek-partition}) and (\ref{eq:ep-partition}) are nicely followed by results of numerical simulations (points) corresponding to the uniform distribution of initial conditions on the constant energy curve.
More precisely, starting from the uniform distribution of initial conditions on the constant energy curve multiple realizations were simulated with the velocity Verlet algorithm \cite{press1992}.
From the generated set of trajectories we calculate $\langle \ek (t) \rangle $ and $\langle \ep (t) \rangle $ as ensemble averages.
Alternatively, using the numerical solution of Eqs.~(\ref{eq:newton1d}) and Eq.~(\ref{eq:period}), one can perform integration  of Eqs.~(\ref{eq:ek-partition-ho}) and (\ref{eq:ep-partition-ho}) for any value of $\nu$.
Both approaches: ensemble averaging (see Figure~\ref{fig:energy-partition}) and numerical integration of Eq.~(\ref{eq:newton1d}) (results not shown) give the same results.
For simplicity, it has been assumed that constant energy curve corresponds to the following initial condition $x(0)=0$ and $v(0)=\sqrt{2}$, i.e., for $m=1$ we have $\ee=1$.

\begin{figure}[!ht]
    \centering
    \includegraphics[width=0.98\columnwidth]{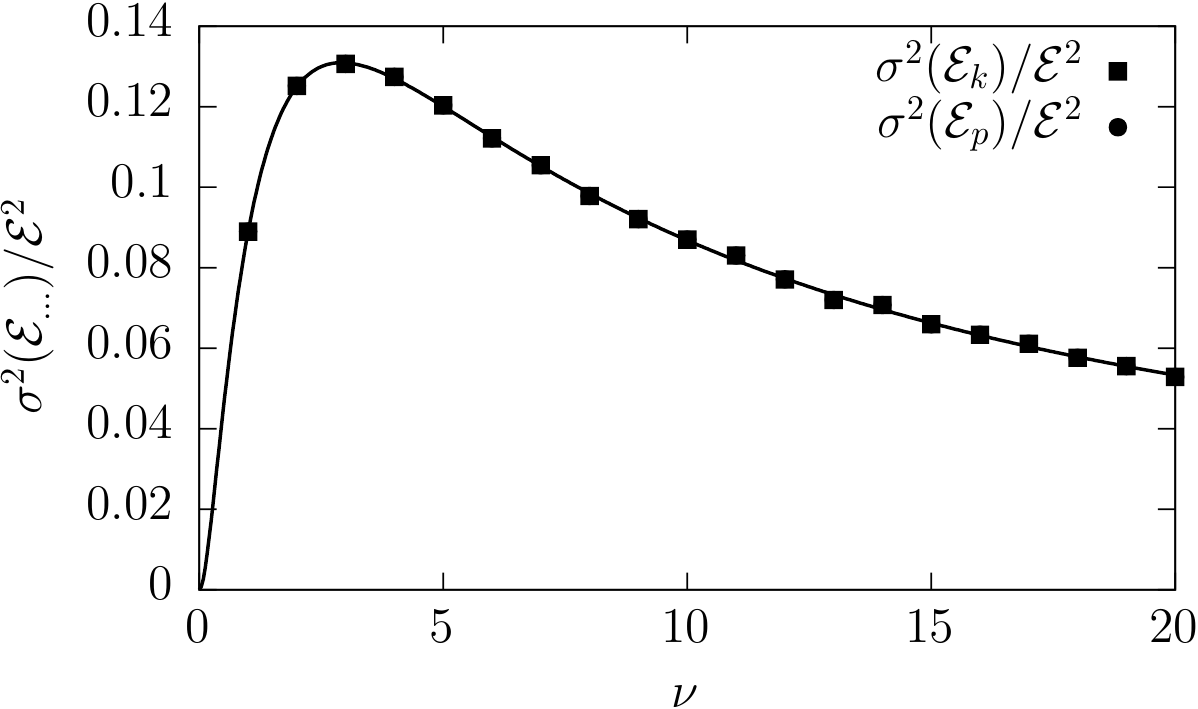} \\
            \caption{Variance of the kinetic $\sigma^2 (\ek) $ (squares) and potential  $\sigma^2 (\ep)^2$ (circles) energies divided by $\ee^2$ as a function of the exponent $\nu$ characterizing the steepness of the potential $V(x)$, see Eq.~(\ref{eq:potential}).
            The solid line depicts the theoretical value given by Eqs.~(\ref{eq:ek-var}) and~(\ref{eq:ep-var}) while superimposed points (squares and circles) correspond to the results of computer simulations.        }
    \label{fig:variance}
\end{figure}

Analogously to Eqs.~(\ref{eq:ek-partition}) and (\ref{eq:ep-partition}), one can calculate variances $\sigma^2(\ek)$ and $\sigma^2(\ep)$
\begin{equation}
 \sigma^2(\ek)=\frac{4\nu^2}{(2+\nu)^2(2+3\nu)}\ee^2
 \label{eq:ek-var}
\end{equation}
and
\begin{equation}
 \sigma^2(\ep)=\frac{4\nu^2}{(2+\nu)^2(2+3\nu)}\ee^2,
 \label{eq:ep-var}
\end{equation}
which are equal.
The equality of variances of kinetic and potential energies emerges as a consequence of the energy conservation.
Fig.~\ref{fig:variance} shows variances of kinetic and potential energies for $1\leqslant \nu \leqslant 20$.
Solid line  given by Eqs.~(\ref{eq:ek-var}) and (\ref{eq:ep-var}) is nicely corroborated by results of numerical simulations corresponding to the uniform distribution of initial conditions on the constant energy curve.
Due to equality of variances of kinetic and potential energies numerically estimated variances are the same,  therefore points (squares and circles) are superimposed.

The variance of kinetic and potential energies, see Eqs.~(\ref{eq:ek-var}) and (\ref{eq:ep-var}) is a non-monotonous function of the exponent $\nu$  characterizing the steepness of the potential $V(x)$, see Eq.~(\ref{eq:potential}). The maximal variance is recorded for $\nu=1+\sqrt{11/3}\approx 2.24$, see Figure~\ref{fig:variance}.
In the limit of $\nu\to\infty$ the variance tends to zero.
Again, it can be intuitively explained by the fact that for $\nu\to\infty$ the potential $V(x)$ transforms into the infinite rectangular potential well and the whole energy is stored in the form of kinetic energy which probability densities are given by $p(\ek)=\delta(\ee-\ek)$ and $p(\ep)=\delta(\ep)$.

\begin{figure}[!ht]
    \centering
    \includegraphics[width=0.97\columnwidth]{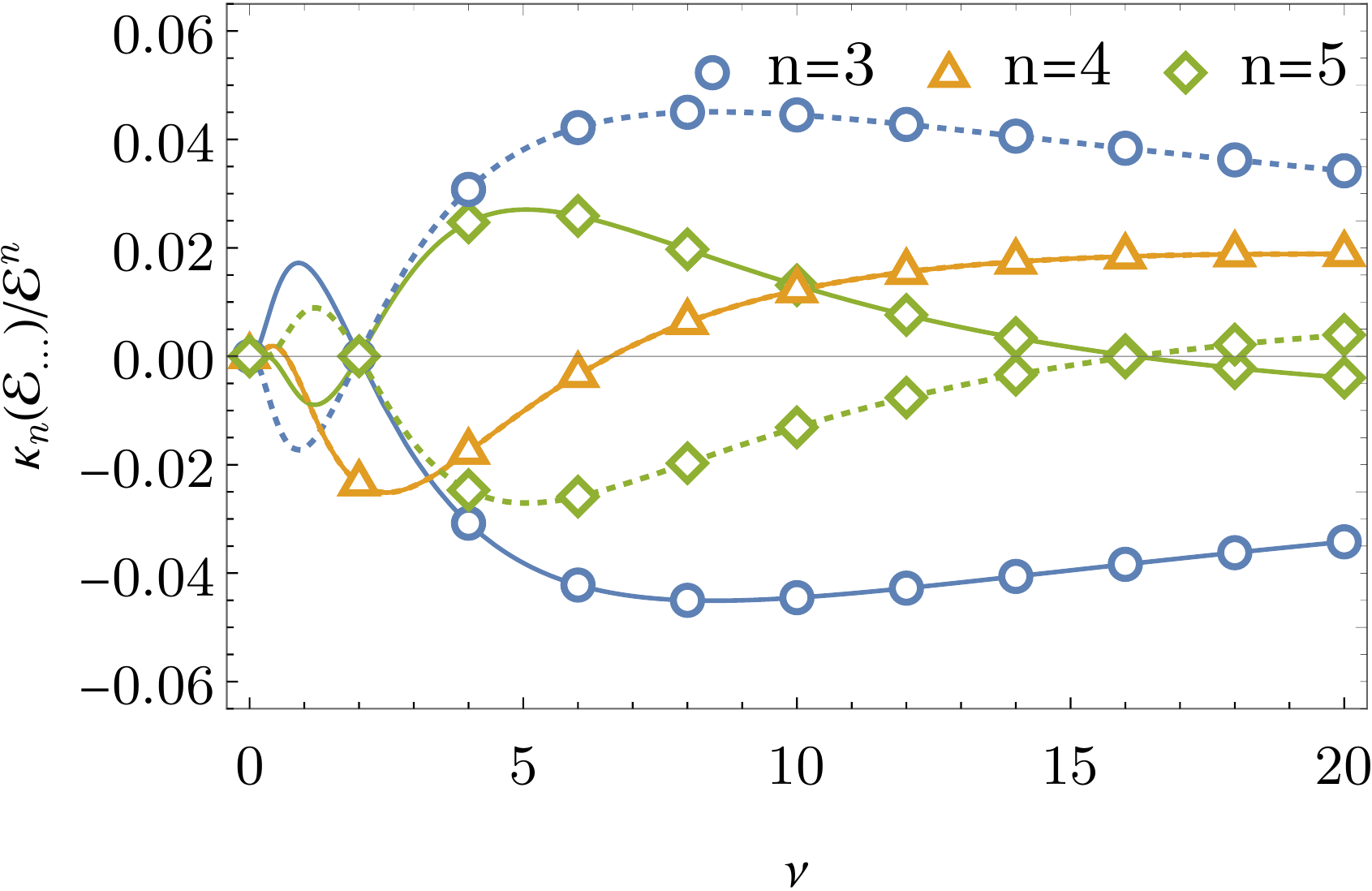} \\
            \caption{Cumulants of the kinetic $\kappa_n(\ek) $ (solid lines) and potential  $\kappa_n(\ep)$ (dotted lines) energies divided by $\ee^n$ as a function of the exponent $\nu$ characterizing the steepness of the potential $V(x)$, see Eq.~(\ref{eq:potential}).
        }
    \label{fig:cumulants}
\end{figure}

Besides the variances, the absolute values of cumulants of kinetic and potential energies, i.e., $\kappa_n(\ek)$, $\kappa_n(\ep)$ for $n\geqslant 2$ are equal.
In particular, for kinetic and potential energies even cumulants are identical while the odd ones (starting from order 3) have an opposite sign, see Figure~\ref{fig:cumulants}. 
The general expressions for the moments (from which the cumulants can be calculated) are given by the formulas
\begin{equation}
    \mu_n(\ek)=\langle \ek^n \rangle= \frac{\Gamma\left( \frac{1}{2}+\frac{1}{\nu} \right) \Gamma\left( n+ \frac{1}{2} \right)}{\Gamma\left( \frac{1}{2}+n+\frac{1}{\nu} \right) \Gamma\left(\frac{1}{2}\right) }\ee^n  
    \label{eq:cumv}
\end{equation}
and
\begin{equation}
    \mu_n(\ep)=\langle \ep^n \rangle= \frac{\Gamma\left( \frac{1}{2}+\frac{1}{\nu} \right)\Gamma\left( n+\frac{1}{\nu} \right)}{\Gamma\left( \frac{1}{2}+n+\frac{1}{\nu} \right) \Gamma\left(\frac{1}{\nu} \right)}\ee^n.
    \label{eq:cumx}
\end{equation}

All cumulants tend to $0$ as the exponent $\nu \to \infty$.
At the same time $\langle \ek^n \rangle$ tends to  $\ee^n$, while $\langle \ep^n \rangle$ goes to zero.
The moment generating functions calculated from Eqs.~(\ref{eq:cumv}) and~(\ref{eq:cumx}) are
$
    M_{\ek}(\xi)={}_1 F_1 \left( \frac{1}{2},\frac{1}{2}+\frac{1}{\nu};\xi \ee \right)
    \label{eq:cumvgen}
$
and
$
    M_{\ep}(\xi)={}_1 F_1 \left( \frac{1}{\nu},\frac{1}{2}+\frac{1}{\nu};\xi \ee \right),
    \label{eq:cumxgen}
$
where ${}_1 F_1(\dots)$ is the Kummer's confluent hypergeometric function.

\section{Discussion}

Thanks to the ergodic theory \cite{walters2000introduction}, it is possible to obtain
formulas for $p(x)$ and $p(v)$ densities, see Eqs.~(\ref{eq:px}) and (\ref{eq:pv}), which can be used to derive exact formulas for the energy partition in (purely) anharmonic, undamped (frictionless), classical oscillators.
Time averaged energies given by Eqs.~(\ref{eq:ek-partition}) -- (\ref{eq:ep-partition})
along with their variances, see Eqs.~(\ref{eq:ek-var}) and (\ref{eq:ep-var}), are the main results of current research.

On the one hand, with the help of time averaging and correspondence principle \cite{liboff2003introductory} it is possible to calculate $p(x)$ and $p(v)$ densities from which energetic characteristics, like the average kinetic or potential energy, of anharmonic oscillators can be obtained.
On the other hand, exact results have been verified by ensemble averaging.
For $V(x) \propto |x|^\nu$ ($\nu>0$) the average potential energy decays with the increasing exponent $\nu$.
At the same time, the average kinetic energy increases with the growing exponent $\nu$.
The equipartition is recorded for $\nu=2$, i.e., for the harmonic oscillator.
Otherwise, for $\nu\neq 2$ the ratio of average energies differs from 1, i.e., it is given by $\langle \ek \rangle / \langle \ep \rangle = \nu/2 $.
The system energy $\ee$ is constant, i.e., $\ep+\ek=\ee=\mathrm{const}$, therefore,
variances of kinetic and potential energies, as well as absolute values of higher cumulants are equal.
Moreover, the variance is a non-monotonous function of the exponent $\nu$.
In the limit of $\nu\to\infty$ the potential well attains the rectangular shape and the particle motion is restricted by two reflecting walls.
The whole energy is stored as the kinetic energy and variances of kinetic and potential energies vanish because energy distributions are given by single Dirac's delta functions.
The presented framework can be extended to any $V(x)$.
In such a case, using Eq.~(\ref{eq:px}) one can easily find the $p(x)$ density.
Contrary to $p(x)$ construction of $p(v)$ distribution is more difficult because the constant energy curve do not need to be convex.

Analogous effect is observed for stochastic underdamped, undamped (frictionless), anharmonic oscillators \cite{mandrysz2019energetics-pre} driven by the Gaussian white noise
\begin{equation}
 m \ddot{x}(t)=-V'(x)+\xi(t),
 \label{eq:stochastic-oscillator}
\end{equation}
where $\xi(t)$ stands for the Gaussian white noise fulfilling $\langle \xi(t) \rangle=0$ and $\langle \xi(t) \xi(s) \rangle=\delta(t-s)$.
Due to absence of damping and action of stochastic force (noise), the system described by Eq.~(\ref{eq:stochastic-oscillator}) is out-of-equilibrium. Asymptotically its average energy $\langle \mathcal{E} \rangle$ grows linearly in time.
For $\nu=2$, $\langle \ek \rangle = \langle \ep \rangle$, while for $\nu\neq 2$ average energies differs. 
Depending on $\nu$: potential ($0<\nu<2$) or kinetic energy ($\nu>2$) energy can dominate, see Ref.~\cite{mandrysz2019energetics-pre}.
Addition of linear damping to Eq.~(\ref{eq:stochastic-oscillator}) 
\begin{equation}
 m \ddot{x}(t)=-V'(x)-\gamma v +\sqrt{2\gamma k_B T/m}\xi(t),
 \label{eq:stochastic-damped-oscillator}
\end{equation}
restores fluctuation dissipation relations and brings the system back to the equilibrium \cite{sekimoto2010stochastic}. 
The stationary state for the model described by Eq.~(\ref{eq:stochastic-damped-oscillator}) not only satisfies the stationary Kramers equation \cite{gardiner2009} but it is also of the Boltzmann--Gibbs type
\begin{equation}
 p(x,v) \propto \exp\left\{ -\frac{1}{k_B T} \left[ \frac{m v^2}{2} + V(x)   \right] \right\}
 \label{eq:bg}
\end{equation}
from which 
\begin{equation}
 \langle \ek \rangle=\frac{k_BT}{2}
 \label{eq:bg-ek}
\end{equation}
and
\begin{equation}
 \langle \ep \rangle=\frac{k_BT}{\nu}.   
 \label{eq:bg-ep}
\end{equation}

On the one hand, averages energies calculated using time averaging, see Eqs.~(\ref{eq:ek-partition}) -- (\ref{eq:ep-partition}), and using Boltzmann--Gibbs  distribution, see Eqs.~(\ref{eq:bg-ek}) -- (\ref{eq:bg-ep}), are different.
The difference between averages can be explained by the energy conservation principle which relates $x$  and $v$ in the studied setup, see Eq.~(\ref{eq:newton1d}), and statistical independence of position and velocity in the Boltzmann--Gibbs distribution.
The Boltzmann--Gibbs distribution~(\ref{eq:bg}) factorizes into position and velocity dependent parts.
Therefore, in the stationary state, despite the functional dependence $\dot{x}(t)=v(t)$, position and velocity are statistically independent.

On the other hand, analogously like in the deterministic model described by Eq.~(\ref{eq:newton1d}) the ratio of average energies reads
$
 \langle \ek \rangle/\langle \ep \rangle=\nu/2.
$
The same results can be obtained by use of the canonical ensemble \cite{huang1963}.
Replacement of the Gaussian white noise with the L\'evy noise moves the system out-of-equilibrium \cite{klages2008,dubkov2008}.
In the regime of linear friction (damping) 
velocity and position are statistically dependent \cite{sokolov2010}, but average energies diverge. Consequently, there is no classical energy equipartition relation \cite{dybiec2017underdamped}.
Nevertheless, systems driven by L\'evy noise require further studies especially in the regime of nonlinear friction \cite{cdegn}, which is capable of bounding average energies.

Within current research we have restricted our analysis to the deterministic motion in the single-well, stable, potentials, i.e., $V(x) \propto |x|^\nu$ with $\nu>0$.
Neverthless, it is possible to condsider motion in unstable potentials.
For instance, noise assisted motion in unstable potentials decaying faster than $1/x$ has been analyzed recently \cite{aghion2019non}, demonstrating the break-down of the standard ergodic hypothesis  and its extension to infinite-ergodic theory \cite{aaronson1997introduction}. 
Moreover, contrary to the stable potentials used within current research, in the case of unstable potentials, e.g., $x^3$, standard statistical quantities of interest can diverge. In such a situation, local characteristics can be analyzed and exploited to obtain effective statistical measures, as it has been shown recently in \cite{vsiler2018diffusing}.


\section*{Acknowledgement}
This research was supported in part by PLGrid Infrastructure.




\def\url#1{}

\end{document}